\documentclass[11pt]{article}

\usepackage[T1]{fontenc}
\usepackage[utf8]{inputenc}
\usepackage{amsmath}
\usepackage{amssymb}
\usepackage{bm}
\usepackage{graphicx}
\usepackage{booktabs}
\usepackage[margin=1in]{geometry}
\usepackage[colorlinks=true,allcolors=blue]{hyperref}
\usepackage{authblk}

\graphicspath{{./}{figures/}}

\newcommand{\yancc}{\texttt{yancc}}

\title{\yancc: A GPU-accelerated, differentiable solver for neoclassical transport in
tokamaks and stellarators}
\author[1]{Rory Conlin}
\author[1]{Matt Landreman}
\affil[1]{Institute for Research in Electronics and Applied Physics,
University of Maryland, College Park, MD 20742, USA}
\date{\today}

\begin{document}

\maketitle

\begin{abstract}
We present \yancc, a new GPU-accelerated solver for the drift kinetic
equation that computes neoclassical transport fluxes, flows, and currents in
tokamaks and stellarators. The drift kinetic equation is challenging to solve
numerically due to strong advection-dominance, recirculating flows, internal
boundary layers, severe anisotropy, and high dimensionality. The code solves
both the full four-dimensional drift kinetic equation (retaining
speed-dependent collisions, energy scattering, and full interspecies
coupling), and the reduced monoenergetic form. The discretization combines a
Maxwell polynomial collocation grid in speed with finite differences in pitch angle
and the flux surface coordinates, using a modified upwind stencil designed to
improve diagonal dominance for multigrid efficiency. The resulting linear
system is solved with a multigrid-preconditioned Krylov method. Built in JAX,
\yancc{} is fully differentiable, enabling gradient-based optimization and
adjoint sensitivity analysis. Benchmarks against MONKES and SFINCS show
agreement within 1\% across a range of collisionalities, geometries, and
multi-species configurations. \yancc{} achieves roughly an order of magnitude
speedup over SFINCS on a per-scan basis while using an order of magnitude less
memory, with runtime remaining nearly flat across the full range of
collisionality. The combination of speed, low memory footprint, and
differentiability makes \yancc{} well suited for integration into stellarator
optimization workflows, uncertainty quantification, and profile prediction.
\end{abstract}

\section{Introduction}
\label{sec:introduction}

Magnetic confinement fusion devices use strong magnetic fields to confine a
high-temperature plasma in toroidal geometry. The magnetic field can be
configured with both toroidal and poloidal components, creating nested flux
surfaces along which charged particles move while radial transport across
surfaces is suppressed. In stellarators, this confining field is generated primarily
through external coils without requiring a toroidal plasma current, providing
inherent steady-state operation and immunity to current-driven disruptions.
However, the three-dimensional magnetic geometry of stellarators makes the
dynamics of confined particles significantly more complex than in axisymmetric
devices such as tokamaks.

Neoclassical transport theory describes the collisional transport of
particles, heat, and momentum in these devices, extending classical transport
to account for the effects of toroidal geometry. The inhomogeneous magnetic
field in a torus causes particle drifts that produce populations of trapped
particles that bounce between magnetic field maxima. These trapped particles,
combined with guiding center drifts of the passing population, yield transport
that exceeds classical predictions and depends sensitively on the
collisionality regime and the structure of the magnetic field. Solving the
drift kinetic equation (DKE) gives the neoclassical particle flux, heat flux,
and parallel flow for each plasma species. From these one computes the radial
electric field via the ambipolarity condition, the bootstrap current, and the
neoclassical conductivity, all essential inputs for transport modeling,
stability analysis, and scenario development.

In stellarators, neoclassical transport differs qualitatively from tokamaks
due to the three-dimensional geometry. The lack of axisymmetry creates
additional ripple-trapped particle populations that can dominate transport at
certain collisionalities. The radial electric field plays a crucial role:
sufficiently strong $E\times B$ rotation suppresses ripple-induced transport,
producing electron-root and ion-root solutions for the ambipolar electric
field with transitions between them as plasma parameters vary. Accurate
computation thus requires a solver that handles the full 3D geometry while
resolving the sharp boundary layers that develop between different classes of
trapped particles at low collisionality.

The numerical solution of the DKE is challenging due to the high resolution
required for the boundary layers, the high-dimensional phase space, strong
advection, anisotropic diffusion, and the need to solve the equation
repeatedly for optimization or parameter scans. To carry out high-resolution solves of
the DKE efficiently, a promising option is a multigrid approach. Multigrid methods
achieve rapid convergence by operating on a hierarchy of discretizations: classical
iterations efficiently damp high-frequency error but stall on low-frequency
components, while multigrid overcomes this by recursively solving coarse-grid
representations where low-frequency fine-grid errors appear high-frequency and
are rapidly damped. Originally developed for elliptic problems
\cite{trottenberg_multigrid_2007, bramble_multigrid_1993,
wesseling_introduction_1995},
multigrid has been extended through algebraic multigrid
\cite{mccormick_algebraic_1987, stuben_review_2001} and geometric
approaches. Advection-diffusion equations present particular difficulties. At
high P\'eclet numbers, strong advective coupling renders standard smoothers
ineffective, motivating upwinded discretizations, line smoothers,
operator-dependent transfer operators, or algebraic multigrid approaches
\cite{de_zeeuw_matrix-dependent_1990, oosterlee_multigrid_1998,
wu_analysis_2006, notay_aggregation-based_2012}. Multigrid has been applied to
the gyrokinetic equation \cite{adams_parallel_2007,
stegmeir_high-performance_2026}, Poisson and Amp\`ere solves in PIC
simulations \cite{chen_multi-dimensional_2015}, and problems in
magnetohydrodynamics \cite{hackbusch_magnetohydrodynamic_1986,
adams_toward_2010}, though its
application to the DKE for neoclassical transport is relatively novel.

A number of codes have been developed to solve the drift kinetic equation.
DKES \cite{hirshman_plasma_1986, van_rij_variational_1989,
shaing_bootstrap_1989} and more recently MONKES \cite{escoto_monkes_2024} use a
spectral Legendre expansion
for the monoenergetic form, eliminating the speed coordinate and thereby
dropping energy scattering and momentum conservation in exchange for
performance. SFINCS \cite{landreman_comparison_2014} solves the full
four-dimensional DKE (dropping only radial coupling) with the full linearized
Fokker-Planck collision operator using a mixture of finite difference and spectral
methods and a sparse direct preconditioner, providing high accuracy across
collisionality regimes at the cost of substantial computational cost. NEO
\cite{belli_eulerian_2009, belli_full_2012, belli_kinetic_2008} adopts a
similar approach and is primarily focused on tokamak geometry, though
extensions to 3D fields also exist \cite{belli_neoclassical_2015}. FORTEC-3D
\cite{satake_non-local_2006, satake_development_2008} is a radially
global code that uses a Monte Carlo approach to solve the DKE, and can capture
full orbit width effects. PIC codes such as Euterpe \cite{kleiber_euterpe_2024,
kuczynski_self-consistent_2024} can also be used to
solve the global drift kinetic equation, though the cost is significantly
higher than 4D codes such as SFINCS, making it impractical for optimization or
large database studies. KNOSOS \cite{velasco_knosos_2020} uses a bounce averaged
approach to
reduce the dimensionality, making it extremely fast, but contains only pitch
angle scattering and is limited to low collisionality regimes, making it
inaccurate when dealing with impurities. NEO-2 \cite{kernbichler_solution_2016,
kernbichler_recent_2008} uses field line
following coordinates and the full linearized collision operator, making it
accurate across a wide range of collisionality, but neglects the effects of the
$\mathbf{E} \times \mathbf{B}$ drift which moves particles across field lines.

Existing neoclassical solvers make unavoidable compromises between speed and
physics fidelity. The monoenergetic approximation sacrifices interspecies
coupling and energy scattering, limiting accuracy for multi-species plasmas at
high collisionality. Full 4D solvers like SFINCS require substantial memory and
scale poorly to GPU hardware due to sparse matrix factorizations with irregular
memory access. We attempt to fill this gap with \yancc{} (``Yet Another
NeoClassical Code''), providing a fast solver for the full DKE with minimal
physics approximations by exploiting GPU parallelism for high-throughput
parameter scans and optimization. Built on JAX \cite{jax2018github}, \yancc{} is
automatically differentiable, enabling gradient-based optimization and
uncertainty quantification workflows that require derivatives of transport
coefficients with respect to input parameters.

\section{Drift Kinetic Equation}
\label{sec:dke}

The drift kinetic equation \cite{hazeltine_recursive_1973} governs neoclassical
transport in toroidal
confinement devices. It has been derived a number of times in the literature
\cite{helander_collisional_2005, hinton_theory_1976}, so we only focus on the
salient points here. We expand the kinetic
equation for small normalized gyroradius $\rho_* = m v_{th} / (|q| B_0 a)$
(here $m$ is the particle mass, $q$ the charge, $v_{th}$ the thermal speed,
$B_0$ the average field strength, and $a$ some characteristic macroscopic
length scale, taken here to be the minor radius) and take the guiding center
distribution function for species $s$ as
\begin{equation}
f_s = f_{1,s} + F_{M,s}
\end{equation}
Where $F_{M,s}$ is a Maxwellian, and we assume that the first order correction
$f_{1,s}/F_{M,s} \sim \rho_s^*$ which is well satisfied in the core of tokamaks
and stellarators. The guiding center distribution function depends on 5
variables, 3 in real space and 2 in velocity space. In real space we
parameterize a toroidal volume by coordinates $(\rho, \theta, \zeta)$ where
$\rho = \sqrt{\psi_N}$ is a radial coordinate which labels flux surfaces, equal
to the square root of the normalized toroidal flux $\psi$ enclosed by the
surface (not to be confused with the normalized gyroradius $\rho_*$), and
$\theta$ and $\zeta$ are general poloidal and toroidal angles on each surface
(we make no assumptions that field lines are straight in these coordinates,
just that the field is tangent to surfaces of constant $\rho$). In these
coordinates the magnetic field can be written as
\begin{equation}
\mathbf{B} = B_\theta \nabla \theta + B_\zeta \nabla \zeta + B_\rho \nabla \rho
= B^\theta \tfrac{\partial \mathbf{r}}{\partial \theta}
+ B^\zeta \tfrac{\partial \mathbf{r}}{\partial \zeta}
\end{equation}
and the Jacobian determinant of the coordinate system is
\begin{equation}
\sqrt{g} = \tfrac{\partial \mathbf{r}}{\partial \rho} \cdot
\tfrac{\partial \mathbf{r}}{\partial \theta} \times
\tfrac{\partial \mathbf{r}}{\partial \zeta}
\end{equation}

In the small $\rho_*$ limit we drop radial coupling, reducing the problem to 4D
($\theta$, $\zeta$, and 2 velocity space coordinates). In velocity space we
choose our coordinates to be normalized speed $x_s = v/v_{th,s}$ and the pitch
angle $\alpha = \arccos(-v_{||}/v)$. In these coordinates, the drift kinetic
equation becomes:
\begin{equation}
\begin{split}
\dot{x}_s \frac{\partial f_{1,s}}{\partial x_s}
+ \dot{\alpha} \frac{\partial f_{1,s}}{\partial \alpha}
+ \dot{\theta} \frac{\partial f_{1,s}}{\partial \theta}
+ \dot{\zeta} \frac{\partial f_{1,s}}{\partial \zeta}
- \sum_{s'} C^\ell_{ss'} + S_s
&= -\mathbf{v}_{\mathrm{drift},s} \cdot \nabla \rho
\Bigg(\frac{\partial F_{M,s}}{\partial \rho} \Bigg)_{W_{0,s}} \\
&\quad + \frac{q_s}{T_s}v_{||}B
\frac{\langle E_{||}B \rangle}{\langle B^2 \rangle} F_{M,s}
\end{split}
\end{equation}
Here and throughout the angle brackets $\langle h \rangle$ denote the flux surface
average of a quantity $h$. The trajectories are given by
\begin{equation}
\dot{\theta} = -\frac{v_{th,s} x_s \cos{\alpha}~ B^\theta}{B}
- \frac{B_\zeta}{B^2\sqrt{g}}E_\rho
\end{equation}
\begin{equation}
\dot{\zeta} = -\frac{v_{th,s} x_s \cos{\alpha} ~ B^\zeta}{B}
+ \frac{B_\theta}{B^2\sqrt{g}}E_\rho
\end{equation}
\begin{equation}
\dot{\alpha} = -\frac{v_{th,s} x_s\sin{\alpha}}{2 B^2}
\Bigg( B^\theta \frac{\partial B}{\partial \theta}
+ B^\zeta \frac{\partial B}{\partial \zeta} \Bigg)
+ \cos{\alpha} \sin{\alpha} \frac{1}{2B^3 \sqrt{g}} E_\rho
\Bigg( B_\zeta \frac{\partial B}{\partial \theta}
- B_\theta \frac{\partial B}{\partial \zeta} \Bigg)
\end{equation}
\begin{equation}
\dot{x}_s = -(1 + \cos^2 \alpha) \frac{x_s}{2B^3 \sqrt{g}} E_\rho
\Bigg( B_\zeta \frac{\partial B}{\partial \theta}
- B_\theta \frac{\partial B}{\partial \zeta} \Bigg)
\end{equation}
Here the radial electric field enters as
$E_\rho = - \partial \Phi_0 / \partial \rho$ with $\Phi_0$ the leading
order potential that is constant on a flux surface,
$v_{th,s} = \sqrt{2T_s/m_s}$ is the thermal speed of species $s$, and $B$ is the
magnetic field magnitude. $S_s$ is a source term that will be described below.

The collision operator $C^\ell_{ss'}$ is the Fokker-Planck-Landau operator
linearized around a Maxwellian:
\begin{align}
C^\ell_{ss'} &= C_{ss'}[f_{1,s}, F_{M,s'}] + C_{ss'}[F_{M,s}, f_{1,s'}] \\
&= C_{L,ss'} + C_{E,ss'} + C_{F,ss'}
\end{align}
$C_L + C_E$ make up the test particle part of the collision operator, with
$C_L$ the Lorentz pitch angle scattering operator between $f_{1,s}$ and
$F_{M,s'}$:
\begin{equation}
C_{L,ss'} = \frac{\nu_{D,ss'}}{2 \sin \alpha} \frac{\partial}{\partial \alpha}
\Bigg[\sin{\alpha} \frac{\partial f_{1,s}}{\partial \alpha}\Bigg]
\end{equation}
and $C_E$ the energy scattering operator between $f_{1,s}$ and $F_{M,s'}$:
\begin{equation}
C_{E,ss'} = \nu_{||,ss'}\Big[ \frac{v^2}{2} \frac{\partial^2 f_{1,s}}{\partial
v^2}
- \frac{v^2}{v_{th,s'}^2} \Big(1 - \frac{m_s}{m_{s'}} \Big) v
\frac{\partial f_{1,s}}{\partial v}\Big]
+ \nu_{D,ss'} v \frac{\partial f_{1,s}}{\partial v}
+ 4\pi \Gamma_{ss'}\frac{m_s}{m_s'} F_{M,s'} f_{1,s}
\end{equation}
$C_F$ is the field-particle scattering operator with Rosenbluth potentials
$H_{s'}$ and $G_{s'}$
\begin{align}
C_{F,ss'} &= \Gamma_{ss'} F_{M,s} \Big[\frac{2v^2}{v_{th,s}^4}
\frac{\partial^2 G_{s'}}{\partial v^2}
- \frac{2v}{v_{th,s}^2} \Big(1 - \frac{m_s}{m_{s'}} \Big)
\frac{\partial H_{s'}}{\partial v}
- \frac{2}{v_{th,s}^2} H_{s'} + 4\pi \frac{m_s}{m_{s'}} f_{1,s'} \Big] \\
\nabla^2_v H_{s'} &= -4 \pi f_{1,s'} \\
\nabla^2_v G_{s'} &= 2 H_{s'}
\end{align}
The collision frequencies and related coefficients are given by
\begin{align}
\nu_{D,ss'} &= \frac{\Gamma_{ss'} n_{s'}}{v^3}
[\mathrm{erf}(v/v_{th,s'}) - \Psi(v/v_{th,s'})] \\
\nu_{||,ss'} &= 2 \frac{\Gamma_{ss'} n_{s'}}{v^3} \Psi(v/v_{th,s'}) \\
\Gamma_{ss'} &= \frac{4\pi q_s^2 q_{s'}^2 \ln \Lambda_{ss'}}
{(4\pi \epsilon_0)^2 m_s^2} \\
\Psi(x) &= \frac{1}{2x^2}\Big[\mathrm{erf}(x)
- \frac{2x}{\sqrt{\pi}} \exp(-x^2) \Big]
\end{align}
where $\ln \Lambda_{ss'}$ is the Coulomb logarithm.

The right hand side drive term is given by
\begin{align}
R_s &= -\mathbf{v}_{\mathrm{drift},s} \cdot \nabla \rho
\Bigg(\frac{\partial F_{M,s}}{\partial \rho} \Bigg)_{W_{0,s}}
+ \frac{q_s}{T_s}v_{||}B \frac{\langle E_{||}B \rangle}{\langle B^2 \rangle}
F_{M,s} \\
&= - (\mathbf{v}_{\mathrm{drift},s} \cdot \nabla\rho)
\Big[\frac{1}{n_s}\frac{\partial n_s}{\partial \rho}
+ \frac{q_s}{T_s}\frac{\partial \Phi_0}{\partial \rho}
+ (x_s^2 - \tfrac{3}{2})\frac{1}{T_s}\frac{\partial T_s}{\partial \rho} \Big]
F_{M,s} \notag \\
&\quad + \frac{q_s}{T_s}v_{||}B
\frac{\langle E_{||}B \rangle}{\langle B^2 \rangle} F_{M,s} \\
&= \Big(\frac{m_s v^2}{q_s} \frac{1 + \cos^2 \alpha}{2B^3}
\mathbf{B} \times \nabla \rho \cdot \nabla B \Big)
\Big[\frac{1}{n_s}\frac{\partial n_s}{\partial \rho}
+ \frac{q_s}{T_s}\frac{\partial \Phi_0}{\partial \rho}
+ (x_s^2 - \tfrac{3}{2})\frac{1}{T_s}\frac{\partial T_s}{\partial \rho} \Big]
F_{M,s} \notag \\
&\quad + \frac{q_s}{T_s}v_{||}B
\frac{\langle E_{||}B \rangle}{\langle B^2 \rangle} F_{M,s}
\end{align}
here the radial derivative of the leading order Maxwellian is
taken at constant leading order energy:
\begin{equation}
W_{0,s} = \frac{m_s v^2}{2} + q_s \Phi_0
\end{equation}
The drive term is commonly split into three linearly independent components:
\begin{equation}
R_s = (A_{1,s} + x_s^2 A_{2,s}) (-\mathbf{v}_{m,s} \cdot \nabla \rho) F_{M,s}
+ A_{3,s} Bv_{||} F_{M,s}
\end{equation}
where the thermodynamic forces are given by
\begin{align}
A_{1,s} &= \frac{1}{n_s}\frac{\partial n_s}{\partial \rho}
- \frac{q_s E_\rho}{T_s}
- \frac{3}{2} \frac{1}{T_s} \frac{\partial T_s}{\partial \rho}, \\
A_{2,s} &= \frac{1}{T_s}\frac{\partial T_s}{\partial \rho}, \\
A_{3,s} &= \frac{q_s}{T_s} \frac{\langle E_\parallel B \rangle}{\langle B^2
\rangle}.
\end{align}
Given solutions to the drift kinetic equation, one can compute neoclassical
fluxes and flows, the most common of which include the per-species surface
average particle flux:
\begin{equation}
\langle \Gamma_s \rangle = \Bigg \langle \int f_{1,s}
\mathbf{v}_\mathrm{drift} \cdot \nabla \rho ~\mathrm{d}^3 \mathbf{v}
\Bigg \rangle
\end{equation}
per-species surface average heat flux:
\begin{equation}
\langle Q_s \rangle = \Bigg \langle \int \frac{m_s v^2}{2} f_{1,s}
\mathbf{v}_\mathrm{drift} \cdot \nabla \rho ~\mathrm{d}^3 \mathbf{v}
\Bigg \rangle
\end{equation}
per-species surface average parallel flow:
\begin{equation}
\langle V_{||,s}B \rangle = \Bigg \langle B \int v_{||} f_{1,s}
\mathbf{v}_\mathrm{drift} \cdot \nabla \rho ~\mathrm{d}^3 \mathbf{v}
\Bigg \rangle
\end{equation}
the radial current:
\begin{equation}
\langle J_\rho \rangle = \sum_s q_s \langle \Gamma_s \rangle
\end{equation}
and the parallel current (including both bootstrap and Ohmic components):
\begin{equation}
\langle J_{||} B \rangle = \sum_s q_s \langle V_{||,s} B \rangle
\end{equation}

In addition to the full 4D problem described above, \yancc{} can also solve a
reduced set of equations commonly called the ``monoenergetic drift kinetic
equation'' which is also solved by DKES \cite{hirshman_plasma_1986} and MONKES
\cite{escoto_monkes_2024}. For the
monoenergetic form we drop all coupling in speed, reducing the problem to 3
dimensions (2 real space coordinates and pitch angle) which also makes it
species independent:
\begin{equation}
\dot{\alpha} \frac{\partial f_{j}}{\partial \alpha}
+ \dot{\theta} \frac{\partial f_{j}}{\partial \theta}
+ \dot{\zeta} \frac{\partial f_{j}}{\partial \zeta}
- \hat{\nu} \mathcal{L} f_{j} = s_j
\end{equation}
The monoenergetic equation uses a modified set of trajectories to ensure
conservation of the magnetic moment and to remove all direct dependence on
velocity:
\begin{equation}
\dot{\theta} = -\frac{ \cos{\alpha}~ B^\theta}{B}
- \frac{B_\zeta}{\langle B^2 \rangle \sqrt{g}}\hat{E}_\rho
\end{equation}
\begin{equation}
\dot{\zeta} = -\frac{\cos{\alpha} ~ B^\zeta}{B}
+ \frac{B_\theta}{\langle B^2 \rangle \sqrt{g}}\hat{E}_\rho
\end{equation}
\begin{equation}
\dot{\alpha} = - \frac{\sin{\alpha}}{2 B^2}
\Bigg( B^\theta \frac{\partial B}{\partial \theta}
+ B^\zeta \frac{\partial B}{\partial \zeta} \Bigg)
\end{equation}
Here the monoenergetic electric field is defined as $\hat{E}_\rho = E_\rho/v$.
In the collision operator we keep only the Lorentz pitch angle scattering with
the monoenergetic collisionality $\hat{\nu} = \sum_{s'} \nu_{D,ss'} / v$. The
right hand side is made up of 3 components corresponding to the three
thermodynamic forces, and the equation is solved once for each right hand side:
\begin{align}
s_1 &= \frac{1 + \cos^2 \alpha}{2B^3} \mathbf{B} \times \nabla \rho \cdot \nabla
B \\
s_2 &= s_1 \\
s_3 &= -\cos{\alpha} ~B
\end{align}
giving the three monoenergetic solutions $f_1, f_2, f_3$. These are then used
to compute the so called monoenergetic transport coefficients
\begin{equation}
D_{ij} = \Bigg\langle \int_0^\pi s_i f_j \sin{\alpha} ~\mathrm{d}\alpha
\Bigg\rangle
\end{equation}

\subsection{Boundary conditions}
\label{sec:bcs}

Using generalized angle like coordinates in real space allows us to use simple
periodic boundary conditions in $(\theta, \zeta)$, as opposed to more
complicated conditions required when using field line following coordinates
such as in KNOSOS \cite{velasco_knosos_2020}. Using the pitch angle $\alpha$
gives symmetric
boundary conditions at $\alpha=0$ and $\alpha=\pi$, requiring
$f(-\alpha) = f(\alpha)$ and $f(\pi -\alpha) = f(\pi + \alpha)$, as opposed to a
regularity type condition that appears when using $\xi = -\cos \alpha$. In
speed, we require that $f \rightarrow 0$ as $x \rightarrow \infty$ which we will
impose with the choice of basis for the speed coordinate in
Section~\ref{sec:discretization}.

\subsection{Properties of the drift kinetic equation}
\label{sec:properties}

Both the forms of the drift kinetic equation have the form of a linear
advection diffusion equation, for which there is a rich history of methods
\cite{morton_numerical_1996}. The prototypical equation for linear advection
diffusion is
\begin{equation}
\mathbf{w} \cdot \nabla f - \nu \nabla^2 f = S
\end{equation}
however, several features of the drift kinetic equation make it more
complicated than the prototypical example. In the prototypical equation, the
problem generally becomes easier for $\nu \gg 1$ as it becomes more Poisson
like, while it becomes singular as $\nu \rightarrow 0$. In the drift kinetic
equation the diffusive term (the collision operator) acts only in velocity
space, so the diffusion is highly anisotropic -- there is zero diffusion in the
real space coordinates. This makes the equation singular in both limits of
collisionality, as $\nu \rightarrow \infty$ we lose coupling in real space. In
practice this is only a minor issue, as the normalized collisionality $\nu^*$
is typically in the range $(10^{-4}, 10^{-2})$ though for impurities it can be
several orders of magnitude larger.

Another important feature of the drift kinetic equation is that all
collisionless characteristics are effectively closed. This is strictly true for
trapped particles, as well as passing particles on rational surfaces. On
irrational surfaces, the trajectories approximately close after a sufficient
number of transits. The periodic boundary conditions in real space combined
with the symmetry/reflecting boundaries in pitch angle means that no
information enters or leaves the domain, which will affect the choice of
smoothing operation in the multigrid scheme, described in
Section~\ref{sec:smoothing}.

One can show \cite{landreman_comparison_2014} that the drift kinetic equation
has a null space of
dimension $2n_s$ where $n_s$ is the number of species consisting of
$f_{1,s} = c_s F_{M,s}$ and $f_{1,s} = c_s v^2 F_{M,s}$ where $c_s$ is an
arbitrary constant. The monoenergetic form has a null space of dimension 1,
corresponding to $f_j = c$ for some constant $c$. To remove this null space, we
impose additional gauge constraints. For the full drift kinetic equation, we
require that all of the density and pressure reside in the leading order
Maxwellian for each species:
\begin{align}
\Bigg \langle \int f_{1,s} ~\mathrm{d}^3 \mathbf{v} \Bigg \rangle &= 0 \\
\Bigg \langle \int v^2 f_{1,s} ~\mathrm{d}^3 \mathbf{v} \Bigg \rangle &= 0
\end{align}
In addition to this gauge freedom, one can also show that the steady state
drift kinetic equation requires an additional source for it to be solvable when
$E_\rho$ is nonzero and not at its ambipolar value (ie $E_\rho$ such that
$J_\rho=0$). To account for this we add additional terms that serve as a source
of particles and heat for each species to ensure solvability for arbitrary
$E_\rho$. For the trajectory models considered here the particle source is not
strictly necessary (the source is always zero at the solution) but it helps by
making the overall system square, and is needed when alternative trajectory
models are considered \cite{landreman_comparison_2014}. The source term for each
species has the form:
\begin{equation}
S_s(x_s) = S_{s,p}(x_s) F_{M,s}(x_s)(x_s^2 - \tfrac{5}{2})
+ S_{s,h}(x_s) F_{M,s}(x_s)(x_s^2 - \tfrac{3}{2})
\end{equation}
where $S_{s,p}$ and $S_{s,h}$ are the free parameters to be solved for.

For the monoenergetic equation, the only gauge freedom that exists is the
choice of the average of $f_j$. One could impose an additional constraint
requiring eg, zero average, but it is simplier to just require that $f$ take a
fixed value at a given point:
\begin{equation}
f_j(\alpha=\pi/2, \theta=0, \zeta=0) = \mathrm{const}
\end{equation}

The final important feature, which also appears in the prototypical equation, is
the presence of internal boundary layers that develop in the solution with a
characteristic width $\sqrt{\nu}$. These layers correspond to the accumulation
of particles near transitions between trapping regions
\cite{hinton_transport_1973, ho_neoclassical_1987}, and
necessitates very high resolution, primarily in the real space and pitch angle
coordinates.

\section{Discretization}
\label{sec:discretization}

Because the location of boundary layers doesn't depend on speed $x_s$, we can
often get away with a relatively coarse discretization in the speed coordinate.
Following SFINCS \cite{landreman_comparison_2014}, we use a collocation approach
based on weighted Maxwell
polynomials, which are orthogonal on $(0, \infty)$ with respect to the weight
function $e^{-x^2}$ \cite{landreman_new_2013}. In practice we usually only need
5--10 points in the speed
coordinate with this discretization.

One could consider using a fully spectral method for the drift kinetic
equation, which has been sucessfully used for the monoenergetic form in DKES
and MONKES, however we found that these did not scale well when applied to the
full 4D problem. Due to the sharp boundary layers that develop at low
collisionality, any method requires a large number of grid points or spectral
collocation points to sufficiently resolve these, so the usual reduction in
resolution possible with spectral methods for smooth solutions isn't possible.
Additionally, spectral methods tend to give dense matrices (though using
Legendre polynomials for pitch angle results in a block pentadiagonal system,
the block size would be $n_s n_x n_\theta n_\zeta$ which can be quite large).

Because of this, in all other coordinates we adopt a finite difference
discretization with a defect correction approach where we use a 2nd order
scheme for the preconditioner and a 4th order scheme for the main operator. We
use uniform grids in $\alpha, \theta, \zeta$, with full index grids in
$\theta,\zeta$ and half index grid in $\alpha$:
\begin{align}
\alpha &= \frac{\pi(2j+1)}{2 n_\alpha} \quad &j \in [0, 1, ..., n_\alpha - 1] \\
\theta_j &= \frac{2\pi j}{n_\theta} \quad &j \in [0, 1, ..., n_\theta-1] \\
\zeta_j &= \frac{2\pi j}{N_{FP}~n_\zeta} \quad &j \in [0, 1, ..., n_\zeta-1]
\end{align}
where $N_{FP}$ is the number of field periods of the underlying equilibrium (a
discrete toroidal symmetry common in stellarators). We use standard centered
differences differences for the collision operator, with the Lorentz pitch angle
scattering operator being explicitly split into first and second derivative
parts, with centered differences applied to each:
\begin{equation}
\mathcal{L} = \frac{1}{\sin \alpha} \frac{\partial}{\partial \alpha}
\sin \alpha \frac{\partial}{\partial \alpha}
= \frac{\partial^2}{\partial \alpha^2}
+ \frac{\cos \alpha}{\sin \alpha} \frac{\partial}{\partial \alpha}
\end{equation}
Using the half index grid in $\alpha$ avoids the singular behavior at
$\alpha=0$ and $\alpha=\pi$. To enforce boundary conditions in $\alpha$ we
mirror $f$ around the endpoints which explicitly enforces
$f(-\alpha) = f(\alpha)$ and $f(\pi - \alpha) = f(\pi + \alpha)$.

For the advective terms we use an upwinded stencil. The ``standard'' 2nd/4th
order upwind scheme uses stencil points $(0,1,2)$ and $(0,1,2,3,4)$ respectively,
where these integers give the relative grid indices for a forward difference, with
$0$ being the point the derivative is being evaluated at, positive numbers are grid
points to the right, negative to the left; indices are reversed about 0 for backward
differences.
In contrast, we use a modified stencil containing the points $(0,1,4)$ (2nd
order) and $(-2, 0, 1, 3, 4)$ (4th order). These wider stencils give a larger
element on the diagonal relative to the off-diagonals, resulting in matrices
with a higher degree of diagonal dominance. We quantify this diagonal dominance with
the ratio $d$:
\begin{equation}
d = \frac{|a_{ii}|}{\sum_{j \neq i} |a_{ij}|}.
\end{equation}
A matrix with $d>1$ is diagonally dominant in the traditional sense. In
Table~\ref{tab:stencil} we compare the stencils and coefficients for our
widened 2nd and 4th order schemes vs the standard upwinded stencils, comparing
the degree of diagonal dominance along with the magnitude of the leading order
error constant. We find that these more diagonally dominant stencils
significantly improve the multigrid smoothing (see Section~\ref{sec:smoothing})
with negligible loss in accuracy.

\begin{table}[htbp]
\centering
\begin{tabular}{llccc}
\toprule
Scheme & Stencil & Coefficients & $d$ & $|\text{leading error}|$ \\
\midrule
standard 2nd order & $(0,1,2)$ &
  $\frac{1}{h}(-\frac{3}{2}, 2, -\frac{1}{2})$ & 0.60 & 0.33 \\
wide 2nd order & $(0,1,4)$ &
  $\frac{1}{h}(-\frac{5}{4}, \frac{4}{3}, -\frac{1}{12})$ & 0.88 & 0.67 \\
standard 4th order & $(0,1,2,3,4)$ &
  $\frac{1}{h}(-\frac{25}{12}, 4, -3, \frac{4}{3}, -\frac{1}{4})$ & 0.24 & 0.20
  \\
wide 4th order & $(-2, 0, 1, 3, 4)$ &
  $\frac{1}{h}(-\frac{1}{15}, -\frac{13}{12}, \frac{4}{3}, -\frac{4}{15},
  \frac{1}{12})$ & 0.62 & 0.20 \\
\bottomrule
\end{tabular}
\caption{Comparison of the widened upwind stencils used in \yancc{} against the
standard upwind stencils, showing the diagonal dominance ratio $d$ and the
magnitude of the leading order error constant. Our widened stencils have a larger
relative magnitude on the diagonal for similar error.}
\label{tab:stencil}
\end{table}

An important requirement when considering discretizations for a multigrid
scheme is to ensure numerical stability not just at the finest grid level but at
all levels. Using upwinded finite differences achieves this by adding artificial
numerical diffusion to avoid small grid scale oscillations that can happen when
under-resolving sharp features. This increased numerical stability is not
without cost though. Because the additional diffusion is proportional to the
grid spacing, coarser grids effectively have higher diffusion, which can make
them a poor approximation to the true fine grid problem, causing convergence
issues for a multigrid scheme. This is a well known problem when using multigrid
for advection diffusion problems \cite{trottenberg_multigrid_2007} and
techniques to mitigate it will be
discussed in Section~\ref{sec:coarse}.

The integrals required for output moments are computed with spectral accuracy in
all coordinates. In $x$ we use the Gaussian quadrature weights associated with
the Maxwell polynomials, while in $\theta,\zeta$ we use standard trapezoidal
quadrature which is spectrally accurate for periodic integrands. In $\alpha$ we
use Fejer type 1 quadrature which is equivalent to Chebyshev quadrature in
$\xi = -\cos \alpha$.

For the field part of the collision operator, there are a number of possible
ways to discretize the Rosenbluth potentials. One could treat them as additional
unknowns and solve a system of equations for $f, G, H$, but we adopt a Green's
function approach to analytically solve the Poisson equations for $G, H$
\cite{mollen_impurities_2015}. The details of this derivation are given in
Appendix~\ref{app:rosenbluth}.

\section{Multigrid method and Krylov solver}
\label{sec:multigrid}

When discretized, the drift kinetic equation becomes a large, sparse linear
system of the form $Af=s$. The matrix $A$ is banded in $\theta$ and $\zeta$ (due
to the finite difference discretization), and dense in $\alpha$ and $x$ (due to
the spectral collocation in $x$ and the integral terms in the collision operator
for $\alpha$). The monoenergetic equation is banded in all three coordinates
$(\alpha, \theta, \zeta)$. Due to the high resolution required to resolve the
trapped passing boundary, the resulting system typically has $\sim 10^6 - 10^7$
degrees of freedom ($\sim 10^5$ is typical for the monoenergetic problem). At
this size for this problem, direct sparse solvers were found to be quite poor,
with scipy's \texttt{spsolve} \cite{virtanen_scipy_2020, demmel_supernodal_1999}
failing to solve even the monoenergetic
problem in a reasonable time.

A more efficient approach is to use a Krylov method like GMRES combined with a
good preconditioner. Existing codes like SFINCS use a sparse LU factorization of
an approximate $A$ as a preconditioner, which tends to work well at low to
moderate collisionality, at the cost of a significant amount of memory (often
100s of GB) to perform the sparse factorization. Additionally, by dropping
certain terms from the preconditioner, it becomes far less effective at high
collisionality, requiring many more Krylov iterations. Finally, sparse matrix
factorizations tend to see minimal speedup on GPU, due to the irregular memory
access \cite{silva_distributed_2014, gale_sparse_2020, claus_graphics_2025}.

We opt for a different approach, using a multigrid preconditioner combined with
an outer Krylov solver. The main elements of a multigrid scheme are a hierarchy
of grids, a smoothing operation, and a coarse grid correction. The basic idea is
as follows:
\begin{enumerate}
\item Starting from some initial guess on a fine grid, apply a smoothing
operation which reduces the high frequency components of the error.
\item Restrict the fine grid problem down to a coarser grid. Because the
smoothing eliminated the high frequency error, the coarse grid problem now
captures the dominant error.
\item Solve the coarse grid problem (possibly approximately) to obtain a
correction.
\item Interpolate this correction back to the fine grid.
\item Repeat.
\end{enumerate}

The key idea is that step 3 can be done recursively, using additional coarser
levels to reduce the problem size until it is small enough to solve directly. In
the ideal case, multigrid methods can achieve a grid independent rate of
convergence, where each pass through the multigrid cycle reduces the error by a
fixed fraction, regardless of resolution, meaning that the number of iterations
required to achieve a given error tolerance is constant as the resolution is
increased (compared to many other methods which require a larger number of
iterations at higher resolution). For advection-diffusion problems obtaining
ideal multigrid convergence is generally difficult, but even in this case it can
be a highly efficient preconditioner.

\subsection{Grid hierarchy}
\label{sec:grid-hierarchy}

The first ingredient in a multigrid scheme is a hierarchy of different grids and
a method to transfer information between them. The resolution on the finest grid
is chosen to properly represent the physics of the problem with our given
discretization. We then must define how the coarse grids are constructed.
Because even the finest grid generally requires very low resolution in $x$
($n_x \sim 5$--$10$) we find little benefit in coarsening the $x$ coordinate, so
we retain the finest $x$ grid across all levels. This is a form of
``semi-coarsening'' where one only coarsens the grid in certain directions. In
the other coordinates $(\alpha, \theta, \zeta)$ we coarsen by a factor of 2 in
each direction each time we go down a grid level, for a total reduction in cost
of roughly 8 per grid level. We find that increasing the coarsening factor in
each coordinate to 3 or even 4 only moderately degrades solver performance.
Semi-coarsening in other coordinates was also considered but was found to be
significantly more expensive in time and memory per iteration (due to the
correspondingly larger coarse grids) without a significant decrease in the
number of iterations.

The final choice in grid hierarchy is when to stop coarsening. For simple
problems one can coarsen extremely far, until the coarsest grid contains only a
single node. In our case we find that using a larger coarse grid improves
performance. Typically we coarsen until the coarsest grid has a few thousand
degrees of freedom at which point we directly solve the coarsest grid system
with a dense LU factorization. Going coarser than this is generally inefficient,
as dense linear algebra on GPU is very fast, as well as the increased overhead
of more grid levels. Using an exact solve on a larger grid also reduces the
effects of the additional numerical diffusion on the coarse grid discussed in
Section~\ref{sec:coarse}.

In addition to the grid hierarchy, we must define operations for moving
information between grids. These are generally referred to as restriction
(moving from fine to coarse) and prolongation (from coarse to fine). There are a
number of methods for restriction/prolongation in use in the literature. A
common approach is to use operator dependent methods,
\cite{de_zeeuw_matrix-dependent_1990} where one
considers the direction of flow and diffusive coupling to determine how to move
between grids. We find in our case that prolongating with simple piecewise
linear interpolation is sufficient, with higher order interpolation or operator
dependent methods increasing the per iteration cost and complexity without a
noticeable decrease in the required number of iterations. For restriction we use
the volume weighted transpose of the prolongation operator.

\subsection{Smoothing}
\label{sec:smoothing}

The next key ingredient for a multigrid scheme is an efficient smoothing
operation. For linear problems, this typically takes the form of a classical
relaxation method which takes the form
\begin{equation}
f_{k+1} = f_k + \omega_s K^{-1} (s - A f_k)
\end{equation}
where $K$ is some approximation of $A$ and $\omega_s$ is a
damping/under-relaxation parameter. For Poisson type problems the standard
choice is $K = \mathrm{diag}(A)$, however for advective or anisotropic diffusion
problems this is generally poor due to strong coupling along characteristics
which is ignored by the diagonal approximation. For advection problems, it is
common to take $K$ as the upper or lower triangle of $A$, or a block triangular
approximation where one keeps full coupling in some coordinates and triangular
coupling in others. For problems with open characteristics and Dirichlet
boundary conditions this can be extremely efficient, as the smoothing operation
takes the known boundary value and sweeps it in along characteristics, reducing
the error at all frequencies across the entire domain. As discussed previously
however, the drift kinetic equation has periodic boundary conditions and a wide
variety of closed characteristics. Because of this, instead of a triangular
approximation of $A$ we instead use a block diagonal approximation for
smoothing. This is significantly cheaper to apply (especially on GPU) and
performs just as well in practice.

When using a block diagonal approximation, the choice of coordinate ordering
determines which matrix elements are on the block diagonal and are thus retained
in the smoothing operation. To avoid having to make an explicit choice, we
consider a family of block diagonal smoothers, which each maintain full coupling
along one coordinate \cite{oosterlee_multigrid_1998, elman_finite_2014}. If $A$
is our original matrix, and $P_j$ is a
permutation matrix that reorders the coordinates to put coordinate
$j \in (x, \alpha, \theta, \zeta)$ on the block diagonal, then we take
$K_j = P^{-1}_j \mathrm{block\_diag}(P_j A P_j^{-1}) P_j$, giving 4 smoothers (3
for the monoenergetic problem) which each preferentially smooth the errors in a
different direction. One full smoother pass consists of applying these 4
smoothers in sequence to a given estimate of the solution.

\begin{figure}[htbp]
\centering
\includegraphics{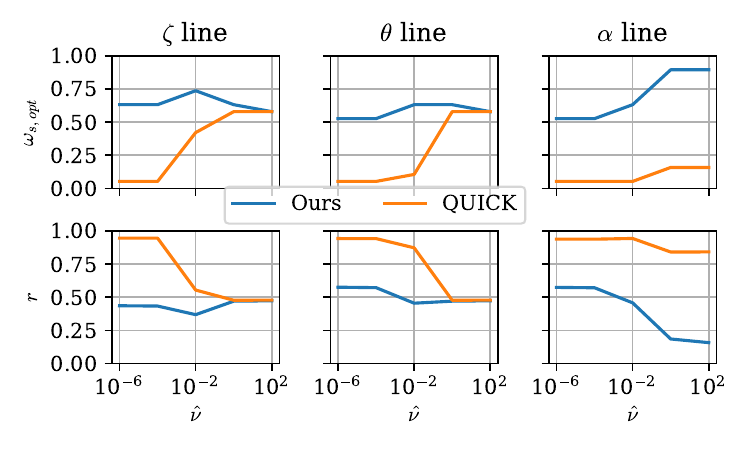}
\caption{Smoothing factor $r$ for the block diagonal smoothers in each
coordinate, comparing the widened upwind stencil used in \yancc{} against the
QUICK scheme on the monoenergetic problem, as a function of collisionality. Using
a wider stencil with a larger diagonal significantly improves the smoothing property.}
\label{fig:smoothing}
\end{figure}

This choice of block diagonal smoothing matrices is aided by our choice of
finite difference stencil (Table~\ref{tab:stencil}), which makes the matrices
more diagonally dominant compared to standard upwinded methods. This larger
element on the diagonal means a block diagonal approximation captures more of
the behavior of $A$, resulting in a more efficient smoother. We can measure the
efficiency of the smoother by considering the operator that multiplies the error
at each smoothing step:
\begin{equation}
f_{k+1} - f_{exact} \equiv e_{k+1} = (I - \omega_s K^{-1} A) e_k
\end{equation}
We can then consider
\begin{equation}
r = \frac{||L(I-\omega_s K^{-1}A) ||}{||L||}
\end{equation}
where $L$ is some approximate high pass filter, which following
\cite{hackbusch_multi-grid_1985} we take
to be the discrete Laplacian on our phase space. This ratio $r$ gives a measure
of how much the smoother is reducing the high frequency errors (lower is
better). In Figure~\ref{fig:smoothing} we compare this ratio for the block
diagonal smoothers in each coordinate for our upwinded stencil vs the ``QUICK''
method (Quadratic Upstream Interpolation for Convective Kinematics, from
\cite{leonard_stable_1979}) on the monoenergetic problem. In both cases we first
optimize the
damping parameter $\omega_s$ to achieve the lowest $r$ possible for each
collisionality considered across a range of different geometries and electric
fields. We see that our method allows a larger damping parameter and gives much
lower smoothing factor, indicating that more of the high frequency error is
eliminated per application, especially at low collisionality for $\theta$,
$\zeta$ and across all collisionalities for $\alpha$. For the full drift kinetic
equation, which spans several orders of magnitude in collisionality, we find that a
fixed damping parameter of $\omega_s = 0.6$ works best.

Because we use a finite difference discretization, the block diagonal itself is
banded for permutations that couple $\theta$ and $\zeta$, and approximately
banded for permutations that couple $\alpha$ (for the smoothing operation we
drop the off band terms corresponding to the integrals in the field part of the
collision operator). This significantly reduces the memory required to store the
smoothers, and due to the larger elements on the diagonal we find that an
unpivoted banded LU factorization is stable, which avoids fill in and irregular
memory access.

\subsection{Coarse grid problem}
\label{sec:coarse}

The standard coarse grid correction takes the form of
\begin{equation}
f_{k+1, fine} = f_{k, fine}
+ \omega_c P A^{-1}_{coarse} R (s_{fine} - A_{fine} f_{k, fine})
\end{equation}
Where $R$ is the restriction operator from fine grid to coarse grid, $P$ is the
prolongation operator from coarse grid to fine grid, and as in the smoothing we
allow for a damping parameter $\omega_c$.

An important consideration in a multigrid scheme is the choice of how to
discretize the coarse grid problem. The two primary methods are direct
discretization, and Galerkin projection. In the direct method, we simply
re-discretize the original problem on the coarser grid, using the same
underlying finite difference method. This has the advantage of being cheap to
implement and apply. Galerkin projection takes
\begin{equation}
A_{coarse} = R A_{fine}P
\end{equation}
This is commonly used in algebraic multigrid schemes and is often recommended
when dealing with advective problems, however it can require more memory and be
difficult to implement in a matrix free manner. The additional memory is due to
the Galerkin projection changing the sparsity structure, meaning that in many
cases $A_{coarse}$ can have more nonzero elements than $A_{fine}$, especially
when combined with semi-coarsening. This also makes coarse grid operations
proportionally more expensive. Additionally, in our case we wish to avoid ever
materializing even a sparse representation of $A_{fine}$ to keep memory usage
low. This is easy when doing direct re-discretization, as applying $A_{fine}$
and $A_{coarse}$ consist of convolutions, small tensor contractions and
elementwise operations, with $A_{coarse}$ being orders of magnitude smaller and
thus cheaper. With a Galerkin projection however, applying $A_{coarse}$ in a
matrix free manner requires interpolating the coarse grid residual back to the
fine grid, applying $A_{fine}$, then restricting. This means that the cost of
the coarse grid operations is as expensive as the fine grid, which
significantly increases the cost of the multigrid cycle. For these reasons we
opt for direct rediscretization which we find works sufficiently well in
practice.

\subsection{Cycle type}
\label{sec:cycle-type}

\begin{figure}[htbp]
\centering
\includegraphics{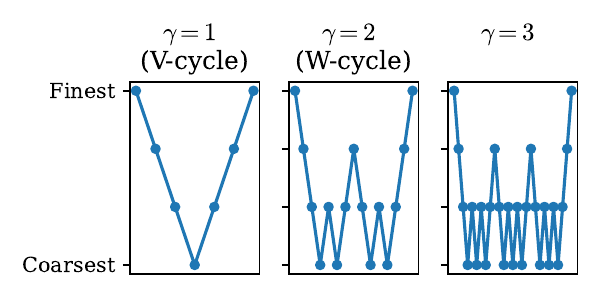}
\caption{Schematic of multigrid cycle types (V-, W-, and higher cycle indices)
across the grid hierarchy.}
\label{fig:cycles}
\end{figure}

The final element of the multgrid scheme to consider is the cycle type,
governed by the ``cycle index''. A cycle index of 1 corresponds to a ``V''
cycle, where we visit each coarser grid once per cycle. Cycle index 2
corresponds to a ``W'' cycle, where we visit each coarser grid twice, etc. These
are shown diagrammatically in Figure~\ref{fig:cycles}. We find that for the
drift kinetic equation cycle index 1 is the most robust. As discussed above, the
coarse grid correction can be unreliable due to additional numerical diffusion,
so higher order cycles that apply this less accurate coarse grid correction more
often tend to slow the solver down. For the monoenergetic equation however, we
find that this is less of an issue and using cycle index 3 tends to be most
efficient.

\subsection{Krylov solver}
\label{sec:krylov}

\begin{figure}[htbp]
\centering
\includegraphics{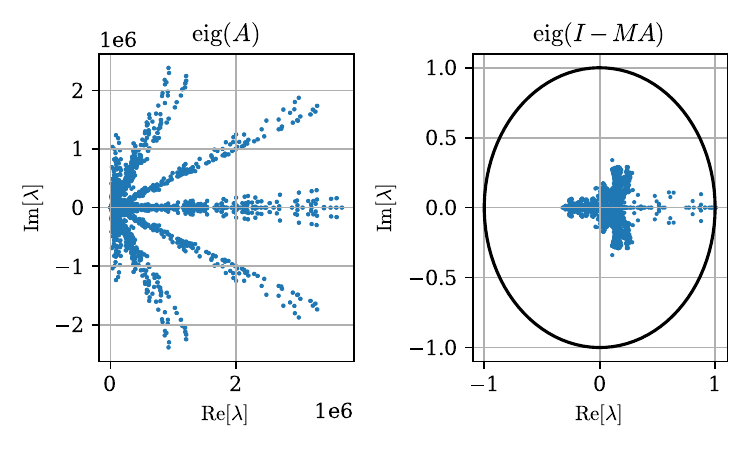}
\caption{Spectrum of the discretized drift kinetic operator $A$ and of the
preconditioned error operator $I-MA$, where $M$ is the multigrid
preconditioner.}
\label{fig:spectrum}
\end{figure}

For many problems a well tuned multigrid method works efficiently as a
standalone solver, however this is often not the case for advection diffusion
equations. This can be seen in Figure~\ref{fig:spectrum}, where we plot the
spectrum of the discretized drift kinetic equation $A$, as well as $I-MA$ where
$M$ is the multigrid preconditioner. This matrix is what multiplies the error at
each step of a pure multigrid cycle. If all of its eigenvalues are within the
unit circle then pure multigrid would converge. We see that in practice, the
multigrid preconditioner clusters the vast majority of the eigenvalues near
zero, which means they are very efficiently damped. However there are several
eigenvalues near the unit circle which are only weakly damped which would slow
convergence, and there can also be outliers that would cause divergence in a
pure multigrid scheme. To handle this, we use multigrid as a preconditioner for
a Krylov solver. We use a right preconditioned GCROT \cite{de1999truncation,
hicken_simplified_2010} with inner GMRES
\cite{saad_gmres_1986, saad_flexible_1993}. The largest memory requirement is
typically storing the Krylov
subspace, so we limit the inner GMRES to 150 iterations before restarting. This
is more than enough for the majority of cases tested, and the outer GCROT loop
efficiently handles restarting by preserving the most important elements of the
Krylov subspace between restarts rather than starting from scratch as in
standard restarted GMRES.

\section{Benchmarks}
\label{sec:benchmarks}

\subsection{W7-X monoenergetic DKE}
\label{sec:bench-w7x}

\begin{figure}[htbp]
\centering
\includegraphics{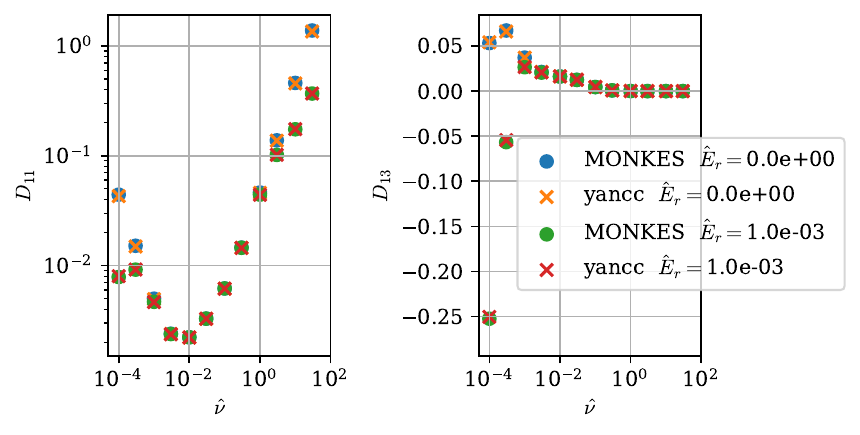}
\caption{Monoenergetic transport coefficients $D_{ij}$ for the W7X-KJM
configuration, comparing \yancc{} against MONKES over a range of collisionality
and electric field.}
\label{fig:mdke-dij}
\end{figure}

As a first example we solve the monoenergetic drift kinetic equation for the
KJM configuration of W7-X over a range of collisionality and electric field. The
results are shown in Figure~\ref{fig:mdke-dij}, comparing \yancc{} vs MONKES
\cite{escoto_monkes_2024}. Both codes are converged to within 1\% across the
full range of
collisionality. The resolutions for each code are given in Table~\ref{tab:w7x}.

\begin{table}[htbp]
\centering
\begin{tabular}{ll}
\toprule
$\rho$ & $0.45$ \\
$\hat{\nu} = \nu_D/v$ & $\in [1 \times 10^{-4}, 3 \times 10^1]$ \\
$\hat{E_r} = E_r/v$ & $[0, 10^{-3}]$ \\
MONKES resolution $(n_l, n_\theta, n_\zeta)$ & $(180, 39, 99)$ \\
\yancc{} resolution $(n_\alpha, n_\theta, n_\zeta)$ & $(201, 31, 81)$ \\
\bottomrule
\end{tabular}
\caption{Parameters and resolutions for the W7-X monoenergetic benchmark.}
\label{tab:w7x}
\end{table}

\begin{figure}[htbp]
\centering
\includegraphics{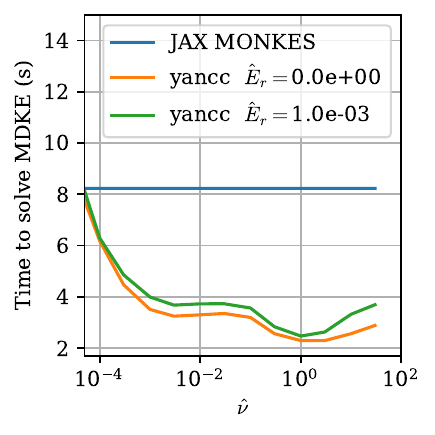}
\caption{Timing comparison between \yancc{} and a JAX re-implementation of the
MONKES algorithm for the monoenergetic problem, as a function of
collisionality.}
\label{fig:mdke-time}
\end{figure}

For a performance comparison, we re-implement the MONKES algorithm in JAX and
compare time and memory required to solve the monoenergetic equation. Both codes
run on the same hardware (Nvidia A100 GPU) to isolate algorithmic differences.
For this problem, \yancc{} required 4GB of memory, while MONKES required only
1.4 GB. This is likely due to the use of a Legendre discretization in MONKES,
and the fact that the monoenergetic coefficients depend only on the first 2
Legendre modes, so very little storage is required. The tradeoff is that MONKES
must invert a large number of dense matrices of size
$(n_\theta n_\zeta) \times (n_\theta n_\zeta)$, which costs
$O(n_\theta^3 n_\zeta^3)$, while \yancc{} requires only operations that scale
linearly with resolution. A timing comparison is shown in
Figure~\ref{fig:mdke-time}, showing \yancc{} is $2\times$--$4\times$ faster for
this problem over most of the collisionality range, though slowing down to
roughly equal at the lowest collisionality. On problems with simpler geometry
that require less resolution in the two flux surface angles, the MONKES
algorithm will likely be faster. As for the discretization and resolution
required, in this case they are broadly similar despite MONKES using a spectral
discretization in all 3 coordinates. This is likely because the solutions are
not sufficiently smooth for spectral convergence to show an advantage. If higher
accuracy is desired then a fully spectral approach would be more efficient
asymptotically, but at this level of accuracy common in optimization and
analysis there is little difference.

\subsection{Single-species density scan}
\label{sec:bench-single}

\begin{figure}[htbp]
\centering
\includegraphics{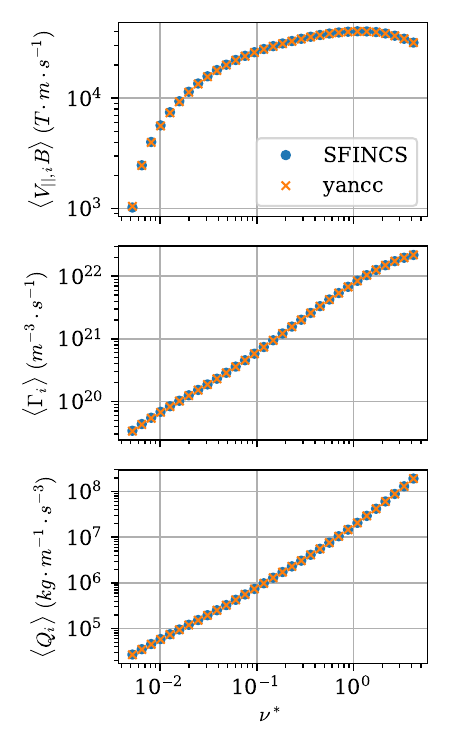}
\caption{Output fluxes (heat flux, particle flux, and parallel flow) for the
single-species NCSX density scan, comparing \yancc{} and SFINCS.}
\label{fig:nhat-fluxes}
\end{figure}

For the full drift kinetic equation, we compare \yancc{} to SFINCS
\cite{landreman_comparison_2014}, here
for a single species scan in density / collisionality using an NCSX like
equilibrium. The parameters are summarized in Table~\ref{tab:ncsx-1s}, and a
comparison of the output fluxes (heat flux $\langle Q \rangle$, particle flux
$\langle \Gamma \rangle$, and parallel flow $\langle V_{||} B \rangle$) is shown
in Figure~\ref{fig:nhat-fluxes}, where again both codes are converged to within
1\%.

\begin{table}[htbp]
\centering
\begin{tabular}{ll}
\toprule
$\rho$ & $0.5$ \\
$E_r = E_\rho/a$ & $-1 ~\mathrm{kV/m}$ \\
$T$ & $800 ~\mathrm{eV}$ \\
$n$ & $\in [1.5 \times 10^{19}, 1.5 \times 10^{22}]~\mathrm{m}^{-3}$ \\
$a/L_T$ & $0.81$ \\
$a/L_n$ & $0.86$ \\
$\nu^*=\frac{R \nu_D}{v_{th} \iota}$ & $5.2\times 10^{-3} - 4.1 \times 10^0$ \\
Species & Hydrogen \\
SFINCS resolution $(n_x, n_\xi, n_\theta, n_\zeta)$ & $(7, 141, 25, 81)$ \\
\yancc{} resolution $(n_x, n_\alpha, n_\theta, n_\zeta)$ & $(7, 121, 43, 65)$ \\
\bottomrule
\end{tabular}
\caption{Parameters and resolutions for the single-species NCSX density scan.}
\label{tab:ncsx-1s}
\end{table}

\begin{figure}[htbp]
\centering
\includegraphics{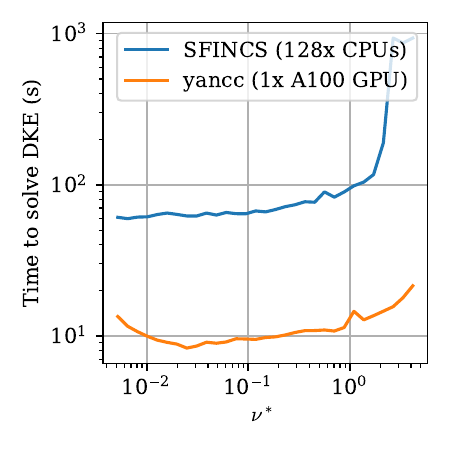}
\caption{Timing comparison between \yancc{} (1x A100 GPU) and SFINCS (128 CPU
cores) for the single-species NCSX density scan, across collisionality.}
\label{fig:nhat-time}
\end{figure}

A timing comparison is shown in Figure~\ref{fig:nhat-time} across the full range
of collisionality. For this case SFINCS was run on 128 CPU cores, while \yancc{}
was run on a single Nvidia A100 GPU, both on the Perlmutter cluster at NERSC and
both using 64 bit floats. We see a speedup of roughly 5x across the full range,
up to nearly 2 orders of magnitude at higher collisionality where SFINCS slows
down significantly. This is likely because the default preconditioner in SFINCS
drops coupling in the speed coordinate, effectively ignoring energy scattering
which becomes important at high collisionality. On the other hand \yancc{}
remains relatively flat, showing only modest slowdown at low or high
collisionality. The times for \yancc{} do not include the cost of just in time
compilation in JAX which must be paid once for a given resolution, after which
the JIT compiled code is re-used across the scan. This is most representative of
real world use where the DKE is solved repeatedly for parameter scans or
optimization tasks.

Due to the different hardware this is not a strictly fair comparison, but to get
a rough estimate we can compare peak theoretical flops. Assuming we could run
SFINCS on GPU with the same overall efficiency as on CPU (unlikely due to the
poor scaling of sparse matrix factorizations on GPU), we would expect a speedup
of roughly 3 going from 128 CPU to 1x A100. We can then see that the speedup
from \yancc{} is due to both being able to run on GPU as well as purely
algorithmic improvements that are independent of hardware. In addition to the
speed improvement, \yancc{} also shows a clear win on memory usage. For these
runs, SFINCS needed more than 50 GB, while \yancc{} needed only 6 GB, which is
especially important for GPU based codes since GPU memory is often a bottleneck.

\subsection{Two-species $E_\rho$ scan}
\label{sec:bench-two}

\begin{figure}[htbp]
\centering
\includegraphics[width=\linewidth]{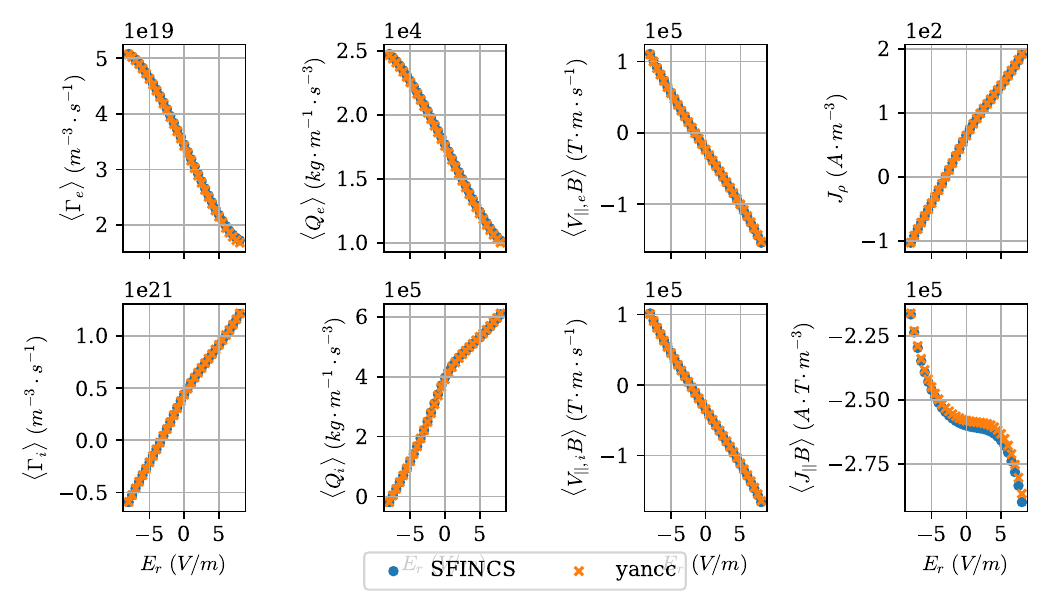}
\caption{Two-species (hydrogen and electron) radial electric field scan for the
NCSX like equilibrium, comparing \yancc{} and SFINCS.}
\label{fig:er-scan}
\end{figure}

To confirm the correctness for the 2 species case with full interspecies
collisions and the ambipolarity condition we consider the same NCSX like
equilibrium but perform a scan in radial electric field for 2 species, electrons
and hydrogen. The parameters are given in Table~\ref{tab:ncsx-2s} and the
results shown in Figure~\ref{fig:er-scan}. As before both codes are converged to
within 1\% for the particle and heat fluxes and parallel flows (particle flux
and heat flux agree to <1\%, the parallel flow is often the hardest to
converge). Composite quantities such as the radial and parallel currents are
also shown. The parallel current shows slightly larger disagreement ($\sim
5\%$), which is to be expected as it is the difference of parallel flows.

As with the comparison to MONKES, we see that \yancc{} requires broadly
similar resolution to SFINCS, though in some cases SFINCS can use slightly
coarser resolution in the flux surface coordinates. This is likely due to the
centered differences that SFINCS uses, which avoids the extra numerical
diffusion caused by upwinding. The cost of this can be stability, and we find
that in some cases \yancc{} can get away with coarser resolution.

\begin{table}[htbp]
\centering
\begin{tabular}{ll}
\toprule
$\rho$ & $0.5$ \\
$E_r = E_\rho / a$ & $\in [-8, 8]~ \mathrm{kV/m}$ \\
$T_H = T_e$ & $800 ~\mathrm{eV}$ \\
$n_H = n_e$ & $1.5 \times 10^{20}~\mathrm{m}^{-3}$ \\
$a/L_{T,H} = a/L_{T,e}$ & $0.81$ \\
$a/L_{n,H} = a/L_{n,e}$ & $0.86$ \\
$\nu^*_H = \frac{R \nu_{D,H}}{v_{th,H} \iota}$ & $5.0 \times 10^{-2}$ \\
$\nu^*_e = \frac{R \nu_{D,e}}{v_{th,e} \iota}$ & $1.3 \times 10^{-1}$ \\
Species & Hydrogen, Electrons \\
SFINCS resolution $(n_x, n_\xi, n_\theta, n_\zeta)$ & $(7, 61, 15, 31)$ \\
\yancc{} resolution $(n_x, n_\alpha, n_\theta, n_\zeta)$ & $(7, 61, 25, 37)$ \\
\bottomrule
\end{tabular}
\caption{Parameters and resolutions for the two-species NCSX radial electric
field scan.}
\label{tab:ncsx-2s}
\end{table}

\section{Conclusion}
\label{sec:conclusion}

We have presented \yancc, a new solver for the drift kinetic equation designed
for GPU acceleration and automatic differentiability. The code solves both the
monoenergetic and full four-dimensional forms of the equation, using a mixed
pseudo-spectral/finite difference discretization with a multigrid preconditioned
Krylov solver. The numerical benchmarks demonstrate agreement with MONKES and
SFINCS to within 1\% across a wide range of collisionalities, electric fields,
and magnetic geometries, for both single-species and multi-species plasmas, and
for both stellarator and tokamak configurations.

In terms of performance, \yancc{} achieves roughly an order of magnitude speedup
over SFINCS on a per-scan basis while using an order of magnitude less memory,
with the gap widening at high collisionality where the SFINCS preconditioner
degrades. The runtime remains nearly flat across a wide range of collisionality,
unlike existing solvers whose iteration counts grow significantly at extremes of
collisionality. This robustness makes \yancc{} practical for production scans
covering the full parameter space of a given device.

The speed and differentiability of \yancc{} open up several applications beyond
standalone neoclassical calculations. In stellarator optimization, neoclassical
transport coefficients must be evaluated many times as the magnetic geometry is
varied, and gradients of these coefficients with respect to geometry parameters
enable efficient gradient-based optimization over the full configuration space.
The low memory footprint and GPU acceleration support the high throughput needed
for uncertainty quantification through Monte Carlo sampling over input parameter
distributions. For integrated modeling and profile prediction, \yancc{} is fast
enough to compute neoclassical coefficients on-the-fly within a transport
solver, replacing the precomputed fit functions or interpolated databases that
are commonly used today. The differentiability also enables adjoint-based
sensitivity analysis, where the derivative of a quantity of interest with
respect to all input parameters can be computed in a single solve.

Several directions for future development are planned. Self-consistent
computation of the ambipolar radial electric field will be added as a built-in
capability, extending beyond the current approach where $E_r$ is specified as an
input. Adaptive mesh refinement in pitch angle and real space is being explored
to efficiently resolve the trapped-passing boundary layers, which can span
several orders of magnitude in width depending on collisionality, without
requiring uniformly fine grids across the full angular domain. Incorporating
corrections for strong radial electric fields and flows near quasisymmetry,
building on the work of \cite{nies_turbulence_2025}, will extend the validity of
the code to regimes
where the conventional $\rho_*$ ordering assumptions begin to break down.
Together, these developments will make \yancc{} a versatile tool for
neoclassical transport calculations across the full range of fusion-relevant
regimes, from conceptual design through discharge analysis.

\subsection{Data Availability}
The source code and data supporting this work is available at 
\url{https://github.com/f0uriest/yancc}

\section{Acknowledgements}
This work was supported by Greg Hammett’s DOE Distinguished Scientist Fellow
award, and by a grant from the Simons Foundation (560651, ML). This research
used resources of the National Energy Research Scientific Computing Center
(NERSC), a Department of Energy User Facility using NERSC award FES-m4505 for
2025-2026. The authors thank Javier Escoto for help with the MONKES benchmark and
for many fruitful discussions about solving the drift kinetic equation, as well
as Greg Hammett for many discussions about subtleties of numerical methods.

\appendix

\section{Rosenbluth potential implementation}
\label{app:rosenbluth}

The Rosenbluth potentials satisfy Poisson type equations:
\begin{equation}
\nabla^2_v H_s = -4 \pi f_{1,s}
\end{equation}
\begin{equation}
\nabla^2_v G_s = 2 H_s
\end{equation}

We can expand these in Legendre polynomials in $\xi = -\cos \alpha$:
\begin{equation}
H_s(\alpha, x_s) = \sum_l P_l(-\cos \alpha) H_{s,l}(x_s)
\end{equation}
\begin{equation}
G_s(\alpha, x_s) = \sum_l P_l(-\cos \alpha) G_{s,l}(x_s)
\end{equation}
And similarly for $f_{1,s}$:
\begin{equation}
f_{1,s}(\alpha, x_s) = \sum_l P_l(-\cos \alpha) f_{1,s,l} (x_s)
\end{equation}

The Legendre harmonics of the potentials are then given by the Greens function:
\begin{equation}
H_{s,l}(x_s) = \frac{4\pi v_{th,s}^2}{2l+1}
\Bigg[\frac{1}{x_s^{l+1}} I_{2,l}(x_s) + x_s^l I_{1,l}(x_s) \Bigg]
\end{equation}
\begin{equation}
G_{s,l}(x_s) = -\frac{4\pi v_{th,s}^4}{4l^2-1}
\Bigg[ {x_s^l} I_{3,l}(x_s) - \frac{2l-1}{2l+3}x_s^{l+2} I_{1,l}(x_s)
- \frac{2l-1}{2l+3} \frac{1}{x_s^{l+1}} I_{4,l}(x_s)
+ \frac{1}{x_s^{l-1}} I_{2,l}(x_s) \Bigg]
\end{equation}
with
\begin{align}
I_{1,l}(x_s) &= \int_{x_s}^\infty z^{-l+1} f_{1,s,l}(z) \mathrm{d} z \\
I_{2,l}(x_s) &= \int_0^{x_s} z^{l+2} f_{1,s,l}(z) \mathrm{d} z \\
I_{3,l}(x_s) &= \int_{x_s}^\infty z^{-l+3} f_{1,s,l}(z) \mathrm{d} z \\
I_{4,l}(x_s) &= \int_0^{x_s} z^{l+4} f_{1,s,l}(z) \mathrm{d} z
\end{align}
We can then express $f_{1,s,l}$ using weighted Maxwell polynomials $L_k(x_s)$:
\begin{equation}
f_{1,s,l}(x_s) = \sum_k f_{1,s,l,k} L_k(x_s) \exp(-x_s^2)
\end{equation}
and define
\begin{align}
I_{1,l,k}(x_s) &= \int_{x_s}^\infty z^{-l+1} L_k(z) \exp(-z^2) \mathrm{d} z \\
I_{2,l,k}(x_s) &= \int_0^{x_s} z^{l+2} L_k(z) \exp(-z^2) \mathrm{d} z \\
I_{3,l,k}(x_s) &= \int_{x_s}^\infty z^{-l+3} L_k(z) \exp(-z^2) \mathrm{d} z \\
I_{4,l,k}(x_s) &= \int_0^{x_s} z^{l+4} L_k(z) \exp(-z^2) \mathrm{d} z
\end{align}
so that
\begin{align}
I_{1,l}(x_s) &= \sum_k f_{1,s,l,k} I_{1,l,k}(x_s) \\
I_{2,l}(x_s) &= \sum_k f_{1,s,l,k} I_{2,l,k}(x_s) \\
I_{3,l}(x_s) &= \sum_k f_{1,s,l,k} I_{3,l,k}(x_s) \\
I_{4,l}(x_s) &= \sum_k f_{1,s,l,k} I_{4,l,k}(x_s)
\end{align}
We can then define
\begin{equation}
H_{s,l,k}(x_s) = \frac{4\pi}{2l+1}
\Bigg[\frac{1}{x_s^{l+1}} I_{2,l,k}(x_s) + x_s^l I_{1,l,k}(x_s) \Bigg]
\end{equation}
\begin{equation}
G_{s,l,k}(x_s) = -\frac{4\pi}{4l^2-1}
\Bigg[ {x_s^l} I_{3,l,k}(x_s) - \frac{2l-1}{2l+3}x_s^{l+2} I_{1,l,k}(x_s)
- \frac{2l-1}{2l+3} \frac{1}{x_s^{l+1}} I_{4,l,k}(x_s)
+ \frac{1}{x_s^{l-1}} I_{2,l,k}(x_s) \Bigg]
\end{equation}
The integrals $I_{1,l,k}$ etc can be cheaply computed as incomplete gamma
functions after a change of variables. Taking $I_{1,l,k}$ as an example, we
first express $L_k$ in the standard monic polynomial form:
\begin{equation}
L_k(x) = \sum_{k'}^k p_{k'} x^{k'}
\end{equation}
then $I_{1,l,k}$ becomes
\begin{equation}
I_{1,l,k}(x_s) = \sum_{k'}^k p_{k'}\int_{x_s}^\infty z^{k'-l+1}
\exp(-z^2) \mathrm{d} z
\end{equation}
Changing variables to $y = z^2$ we obtain
\begin{align}
I_{1,l,k}(x_s) &= \frac{1}{2}\sum_{k'}^k p_{k'}
\int_{x^2_s}^\infty y^{(k'-l)/2} \exp(-y) \mathrm{d} y \\
&= \frac{1}{2}\sum_{k'}^k p_{k'} \Gamma(k'/2-l/2 + 1, x_s^2)
\end{align}
where $\Gamma(s,x)$ is the upper incomplete gamma function. The other integrals
can be transformed similarly, with $I_2$ and $I_4$ giving lower incomplete gamma
functions $\gamma(s,x)$. To avoid numerical cancellation due to the large
magnitudes involved, we compute the gamma functions in logarithmic form and
evaluate the integral using a weighted log-sum-exp operation.

The distribution function $f_s$ is stored on a grid in $(x, \alpha)$ as
$f_{s,x,\alpha}$, and the greens functions $H_{s,l,k}$ and $G_{s,l,k}$ are
stored on the same speed grid, for each cross species interaction as
$H_{s,s',x',l,k}$ and $G_{s,s',x',l,k}$ (where $x' = x_s v_{th,s}/v_{th,s'}$,
the potential for species $s$ evaluated on the speed grid for species $s'$). The
Rosenbluth potentials can then be evaluated as simple tensor contractions:
\begin{equation}
H_{s, s',x',\alpha'} = \sum_{l,k,x,\alpha} T^{-1}_{l,\alpha'} H_{s,s',x',l,k}
V_{k,x}T_{l,\alpha}f_{s,x,\alpha}
\end{equation}
\begin{equation}
G_{s, s',x',\alpha'} = \sum_{l,k,x,\alpha} T^{-1}_{l,\alpha'} G_{s,s',x',l,k}
V_{k,x}T_{l,\alpha}f_{s,x,\alpha}
\end{equation}
where $T_{l,\alpha}$ is the change of basis matrix transforming between pitch
angle $\alpha$ and Legendre index $l$ and $V_{k,x}$ is the change of basis
matrix between speed $x$ and Maxwell polynomial index $k$. Derivatives of the
potentials with respect to speed required for the collision operator are
obtained analytically and precomputed/applied in a similar fashion. The cost of
these tensor operations is quite low, as the resolution in $x$ is fairly coarse
($n_x \sim 5$--$10$), and generally only a small number of Legendre modes are
required to represent the potentials (generally $l_{max} \sim 4$--$6$).

\bibliographystyle{plain}
\bibliography{yancc}

\end{document}